\input harvmac

\def\RR{\hbox{R\kern -.65em {\rm R}}}

\def\lroverarrow#1{\raise4.2truept\hbox{$\displaystyle
\leftrightarrow\atop\displaystyle#1$}}
\def\underarrow#1{\vbox{\ialign{##\crcr$\hfil\displaystyle
 {#1}\hfil$\crcr\noalign{\kern1pt\nointerlineskip}$\longrightarrow$\crcr}}}

%

\font\teneurm=eurm10 \font\seveneurm=eurm7 \font\fiveeurm=eurm5
\newfam\eurmfam
\textfont\eurmfam=\teneurm \scriptfont\eurmfam=\seveneurm
\scriptscriptfont\eurmfam=\fiveeurm

 \font\teneusm=eusm10 \font\seveneusm=eusm7 \font\fiveeusm=eusm5
\newfam\eusmfam
\textfont\eusmfam=\teneusm \scriptfont\eusmfam=\seveneusm
\scriptscriptfont\eusmfam=\fiveeusm

\font\tencmmib=cmmib10 \skewchar\tencmmib='177
\font\sevencmmib=cmmib7 \skewchar\sevencmmib='177
\font\fivecmmib=cmmib5 \skewchar\fivecmmib='177
\newfam\cmmibfam
\textfont\cmmibfam=\tencmmib \scriptfont\cmmibfam=\sevencmmib
\scriptscriptfont\cmmibfam=\fivecmmib


%

%

%
\def\pt{\partial}

\def\bar{\overline}

\def\half{{1 \over 2}}

\def\v{{\rm v}}
\def\x{{\rm x}}
\def\p{{\rm p}}

\def\CD{{\cal D}}

\def\CS{{\cal S}}

\overfullrule=0pt

\Title{} {\vbox{\centerline{Spin Path Integral And Quantum
Mechanics}
\bigskip \centerline{In Rotating Reference Of Frame}}}
\bigskip \centerline{Tong Chern$^\dagger$, Yue Yu, Wu Ning}
\bigskip \centerline{\it $^\dagger$ tongchen@mail.ihep.ac.cn}

\bigskip

\smallskip
\smallskip
\input amssym.tex

\noindent
In the present paper, we develop a path integral formalism of the
quantum mechanics in a rotating reference of frame, and apply this
description to Rabi oscillation, which is important for realizing
the quantum qubit, and to the explanations of various experiments
related to Sagnac effect. In particularly, we propose a path
integral description to spin degrees of freedom, connect it to the
Schwinger bosons realization of angular momentum, and then apply
this description to path integral quantization of spin in the
rotating frame. \Date{}

\newsec{Introduction}
\nref\cavityQED{M. Brune, F. Schmidt-Kaler, A. Maali, J. Dreyer, E.
Hagley, J. M. Raimond, and S. Haroche, ¡°Quantum Rabi Oscillation: A
Direct Test of Field Quantization in a Cavity¡±, Phys. Rev. Lett.
76(1996)1800¨C1803} \nref\quantumcomputer{John M. Martinis, S. Nam,
and J. Aumentado, C. Urbina, ¡°Rabi Oscillations in a Large
Josephson-Junction Qubit¡±, Phys. Rev. Lett. 89(2002)117901; Xiaoqin
Li, Yanwen Wu, Duncan Steel, D. Gammon, T. H. Stievater, D. S.
Katzer, D. Park, C. Piermarocchi, L. J. Sham, ¡°An All-Optical
Quantum Gate in a Semiconductor Quantum Dot¡±, Science
301(2003)5634; J. Johansson, S. Saito, T. Meno, H. Nakano, M. Ueda,
K. Semba, and H. Takayanagi, ¡°Vacuum Rabi Oscillations in a
Macroscopic Superconducting Qubit LC Oscillator System¡±, Phys. Rev.
Lett. 96(2006)127006}\nref\coptics{K. Y. Bliokh, Y. Gorodetski, V.
Kleiner, E. Hasman, ¡°Coriolis Effect in Optics Unified Geometric
Phase and Spin-Hall Effect¡±, Phys. Rev. Lett.
101(208)030404}\nref\colella{A. W. Overhauser and R. Colella,
¡°Experimental Test of Gravitationally Induced Quantum
Interference¡±; R. Colella, A. W. Overhauser and S. A. Werner,
¡°Observation of Gravitationally Induced Quantum Interference¡±}
\nref\sagnac{M. G. Sagnac, C. R. Acad. Sci. 157, 708, (1913)1410; U.
Bonse and M. Hart, Appl. Phys. Lett. 6(1965)155} \nref\sagnacnew{F.
Riehle, Th. Kisters, A. Witte, and J. Helmcke, Ch. J. Bord¨¦,
¡°Optical Ramsey spectroscopy in a rotating frame: Sagnac effect in
a matter-wave interferometer¡±, Phys. Rev. Lett.
67(1991)177¨C180}\nref\neutron{H. Rauch, W. Treimer, and U. Bonse,
Phys. Lett. 47A(1974)425; L. R. Page, ¡°Effect of Earth's Rotation
in Neutron Interferometry¡±, Phys. Rev. Lett. 35(1975)543; J.
Anandan, ¡°Gravitational and Rotational Effects in Quantum
Interference¡±; L. Stodolsky, ¡°Matter and Light Wave Interferometry
in Gravitational Fields¡±, Phys. Rev. D 15(1977)1448; S. A. Werner,
J. L. Staudenmann, R. Colella, ¡°Effect of Earth's Rotation on The
Quantum Mechanical Phase of The Neutron¡±, Phys. Rev. Lett.
42(1979)1103; B. Mashhoon, ¡°Neutron interferometry in a rotating
frame of reference¡±, Phys. Rev. Lett. 61(1988)2639¨C2642}
\nref\etspin{B. Mashhoon, R. Neutze, M. Hannam and G. E. Stedman,
¡°Observable Frequency Shifts via Spin-Rotation Coupling¡±, [arXiv:
gr-qc/9808077]}\nref\rabi{I. I. Rabi, N. F. Ramsey, J. Schwinger,
¡°Use of Rotating Coordinates in Magnetic Resonance Problems¡±, Rev.
Mod. Phys. 26(1954)167} \nref\xgw{Xiao-Gang Wen, ¡°Quantum Field
Theory of Many-Body Systems -- from the origin of sound to an origin
of light and electrons¡±, Oxford University Press, 2004}
\nref\schwinger{J. Schwinger, ¡°Quantum Mechanics Symbolism of
Atomic Measurement¡±, Springer, 2001}

The Quantum mechanics in a rotating reference of frame has many
important applications, for examples, to Rabi oscillation which is
crucial for cavity quantum electro-dynamics \cavityQED\ and for
designing the qubit circuit of a scalable quantum computer (e.g.
\quantumcomputer\ ), to the unified geometric phase and spin-Hall
effect in optics by analyzing the Coriolis effect \coptics\ .

The main purpose of the present paper is to develop a path integral
description of the quantum mechanics in rotating frame. We
investigate a charged particle in the rotating frame with a uniform
external magnetic field applied to it, and use the path integral
description to explain various related experiments, e.g. the sagnac
effect \sagnac\ due to the coupling between the orbital angular
momentum of the particle and the rotation of the reference frame
(see \sagnacnew\ for some important applications of this effect),
and the spin-rotation coupling analog of this effect, in Neutron
interference \neutron\ (see \etspin\ for the experimental
observation of the phase shift via spin-rotation coupling), to solve
Rabi oscillation problem which is traditionally solved in
Hamiltonian formulation \rabi\ .

To achieve these purposes, we present the quantum mechanical theory
of spin degrees of freedom in rotating frame by using the spin path
integral description (see e.g. \xgw\ ) and connecting it to the
Schwinger bosons realization of the algebra of angular momentums
$\vec{J}$. Combing with path integral description to coordinate
variables, we can develop the full path integral theory of a point
particle in a rotating frame. We can then derive the phase factor
due to the spin-rotation coupling and its orbital angular momentum
counterpart (the sagnac effect), give a path integral solution for
Rabbi oscillation problem, and extend the Coriolis force from
classical mechanics to quantum mechanics, by using the quantum
action principle \schwinger\ .

\newsec{The Path Intergal Formalism of Quantum Mechanics In Rotating Frame}

\bigskip\noindent{\it{Spin Path Integral And Schwinger Bosons}}

To describe the spin degrees of freedom of a non-relativistic spin
$s={n\over 2}$ particle with mass $m$ and electric charge $e$, we
introduce the action \eqn\spinaction{I=\int dt
\left[i\phi^{\dagger}{d\over
dt}\phi-\lambda(\phi^{\dagger}\phi-n)\right],} where $\phi$ is a two
components bosonic variable, $\phi^{\dagger}=(\matrix{\bar{\phi^1},
\bar{\phi^2}})$ is its Hermitian conjugation, and $\lambda$ is a
Lagrangin multipler. \spinaction\ has a $U(2)$ symmetry which acts
as $\phi\rightarrow U\phi$, where $U\in U(2)$. To realize the
$SU(2)$ symmetry of spin, one should modulo out a $U(1)$ factor,
which acts as $\phi\rightarrow e^{-i\theta}\phi,
\bar{\phi}\rightarrow e^{i\theta}\bar{\phi}$, by requiring all the
physical observable being $U(1)$ invariant. The conservation charge
of this $U(1)$ transformation is $N(\phi)=\phi^{\dagger}\phi$ which
is itself a physical observable. Other fundamental physical
observable\foot{A little thought show us that, an arbitrary physical
observable can be written as the function of $\vec{S}(\phi)$ and
$N(\phi)$.} are the conservation charges
$\vec{S}(\phi)=\half\phi^{\dagger}\vec{\sigma}\phi$ of the $SU(2)$
symmetry, where $\vec{\sigma}$ are the three Pauli matrixes. In the
quantum theory $\vec{S}(\phi)$ realize the $SU(2)$ symmetry algebra,
which we identify with the spin, thus $\phi$ is a Pauli spinor. In
the path integral quantization, $\vec{S}(\phi)$ can be inserted in
the path integral
$\int{\cal{D}}\phi\CD\bar{\phi}{\cal{D}}\lambda\exp(iI/\hbar)$.

To see the connection between above spin path integral and the
$SU(2)$ algebra more clearly, we now perform the canonical
quantization procedure. Clearly, $i\phi^{\dagger}$ is the canonical
momentum of $\phi$, and the Hamiltonian for the free spin is
trivial. $\phi$ is now realized as the Schwinger bosons with the
commutators $[\phi^{\alpha},
\bar{\phi^\beta}]=\hbar\delta_{\alpha\beta}$, where
$\alpha,\beta=1,2$ are the indexes of $\phi$. The constraint
equation associated with $\lambda$ will restrict us to the Hilbert
space of $n$ Schwinger bosons
$(\phi^{\dagger}\phi-n)|\psi\rangle=0$, where $|\psi\rangle$ is an
arbitrary state of the Hilbert space ${\cal{H}}_s$ of the spin,
$|\psi\rangle\in {\cal{H}}_s$. The basis of ${\cal{H}}_s$ can be
constructed by acting $n$ creation bosons $\bar{\phi^{\alpha}}$ on
the Fock vacuum\foot{In the full Fock space, one can also construct
the spin coherent state $|z\rangle$ by defining it as the
eigenstates of $\phi$, with
$\phi^{\alpha}|z\rangle=z^{\alpha}|z\rangle$. The mean value of the
spin operator $\vec{S}$ on the spin coherent states is $\langle
z|\vec{S}|z\rangle=\half z^{\dagger}\vec{\sigma}z$.},
$|\alpha_1\alpha_2...\alpha_n\rangle
=\bar{\phi^{\alpha_1}}\bar{\phi^{\alpha_2}}...\bar{\phi^{\alpha_n}}|0\rangle$,
where $|0\rangle$ is the Fock vacuum which satisfy
$\phi^{\alpha}|0\rangle=0$. The spin operators $\vec{S}$ are now
realized as \eqn\ssr{\vec{S}=\half \phi^{\dagger}\vec{\sigma}\phi.}
The action of $\vec{S}$ on ${\cal{H}}_s$ gives out $s={n\over 2}$
representation.

\bigskip\noindent{\it{Spin In The Rotating Reference of Frame}}

We now turn to the non-inertial reference of frame $\CS$, which is
rotating with angular velocity $\vec{\omega}(t)$. Firstly, we note
that in terms of the variables in $\CS$, the time changing of spinor
$\phi(t)$ should be \eqn\rotatingderivative{{d\phi\over
dt}-i\half\vec{\omega}(t)\cdot \vec{\sigma}\phi\ ,} which is the
generalization of the ordinary ${d\vec{\v}\over
dt}+\vec{\omega}(t)\times\vec{\v}$ for the time changing--in terms
of variables of $\CS$--of vector $\vec{\v}(t)$. This generalization
is motivated by noticing that under an infinitesimal rotation
$(1-i\delta\vec{\theta}\cdot\vec{J})$, a vector $\vec{\v}$
transforms as $\vec{\v}\rightarrow
\vec{\v}-i(\delta\vec{\theta}\cdot\vec{L})\vec{\v}=\vec{\v}+\delta{\vec{\theta}\times
\vec{\v}}$, and a spinor $\phi$ transforms as $\phi\rightarrow
\phi-i\half\delta\vec{\theta}\cdot \vec{\sigma}\phi$.

Thus, in frame $\CS$, the action for the spin is given by replacing
the ${d\phi\over dt}$ of \spinaction\ with ${d\phi\over
dt}-i\half\vec{\omega}(t)\cdot \vec{\sigma}\phi$
 \eqn\spinraction
{\eqalign{I_{\CS}&=\int dt\left[i\phi^{\dagger}\left({d\over
dt}-i\half\vec{\omega}(t)\cdot
\vec{\sigma}\right)\phi-\lambda(\phi^{\dagger}\phi-n)\right]\cr
&=\int dt\left[i\phi^{\dagger}{d\over dt}\phi+\vec{\omega}(t)\cdot
\half\phi^{\dagger}\vec{\sigma}\phi-\lambda(\phi^{\dagger}\phi-n)\right].}}
In terms of the path integral quantization, spin in the rotating
frame is described as an insertion
$\vec{S}(\phi)=\half\phi^{\dagger}\vec{\sigma}\phi$ in $\int
\CD\phi\CD\bar{\phi}\CD\lambda\exp(iI_{\CS}/\hbar)$.

To compare with the Hamiltonian formalism, one canonically quantize
\spinraction. The Hamiltonian $H_{\CS}$ can be easily got\foot{The
quantization relation of the Schwinger bosons, the constraint
equation and the spin Hilbert space ${\cal{H}}_s$ are all the same
as mentioned in above subsection.},
\eqn\Hh{H_{\CS}=-\vec{\omega}(t)\cdot\vec{S}.} This result is
exactly identical with \rabi\sagnac.

In the static frame, one should insert $\exp(ig{\mu\over \hbar}\int
dt \vec{B}(t)\cdot \vec{S}(\phi))$ into the path integration to
account for the spin-magnetism coupling, where $\mu=e/2mc$, and $g$
is the g factor of the considered particle, and $\vec{B}(t)$ is a
uniform magnetic field. The total effect is to change the action
from $I$ for the free spin to $I_B$ for the spin in $\vec{B}$ field,
where $I_B$ is given as \eqn\actionb{I_B=\int
dt{\left[i\phi^{\dagger}{d\over
dt}\phi-\lambda(\phi^{\dagger}\phi-n)+g\mu\vec{B}(t)\cdot
\vec{S}(\phi)\right].}}

We now generalize the consideration to the rotating frame. For
simplicity, we assume that the angular velocity $\vec{\omega}$ is
time independent, and use $\vec{B}(t)$ to denote the magnetic field
in the rotating frame $\CS$. To write down the explicit form of the
magnetic field in the static inertial frame, we first decompose
$\vec{B}(t)$ into the parallel part $\vec{B}_{\parallel}(t)$--which
is parallel to $\vec{\omega}$--and the perpendicular part
$\vec{B}_{\perp}(t)$--which is perpendicular to $\vec{\omega}$. The
magnetic field in the static frame can then be written as
$\vec{B}_{iner}(t)=\vec{B}_{\parallel}(t)+\vec{B}^{+}_{\perp}(t)e^{i\omega
t}+\vec{B}^{-}_{\perp}(t)e^{-i\omega t}$, where the additional
factors $e^{i\omega t}$ and $e^{-i\omega t}$ are due to the rotation
of $\CS$. \foot{To write out the expressions of
$\vec{B}^{+}_{\perp}$ and $\vec{B}^{-}_{\perp}$ explicitly, one can
coordinates the plane perpendicular to $\vec{\omega}$ as $x-y$
plane, then
$\vec{B}^{+}_{\perp}=\vec{B}^x_{\perp}+i\vec{B}^y_{\perp}$ and
$\vec{B}^{-}_{\perp}=\vec{B}^x_{\perp}-i\vec{B}^y_{\perp}$.}  Now,
the action \spinraction\ and action \actionb\ should be combined
into the action $I_{B\CS }$,
\eqn\spinractionb{\eqalign{I_{B\CS}=\int dt
\left[i\phi^{\dagger}{d\over
dt}\phi+(\vec{\omega}+g\mu\vec{B}(t))\cdot
\vec{S}(\phi)-\lambda(\phi^{\dagger}\phi-n)\right].}} In terms of
the canonical quantization, the Hamiltonian in $\CS$ is $H_{B\CS }$,
with \eqn\hHrB{H_{B\CS}=-g\mu\vec{B}(t)\cdot
\vec{S}-\vec{\omega}\cdot \vec{S},} in agreement with \rabi.

\bigskip\noindent{\it{Include Position Variables}}

We now include the position variables $\vec{\x}$ of the particle in
the rotating frame $\CS$. The spatial part $I_{\x}$ of the total
action is given as \eqn\xraction{\eqalign{I_{\x}&=\int dt\left[\half
m\left({d\vec{\x}\over
dt}+\vec{\omega}\times\vec{\x}\right)^2-V(\vec{\x})+e(\half\vec{B}\times\vec{\x})\cdot
({d\vec{\x}\over dt}+\vec{\omega}\times {\vec{\x}})/c\right]\cr
&=\int dt\left[\half m\left({d\vec{\x}\over
dt}\right)^2+m(\vec{\omega}+\mu\vec{B})\cdot(\vec{\x}\times
{d\vec{\x}\over dt})\right]\cr &-\int dt \left[V(\vec{\x})-\half
m(\vec{\omega}\times\vec{\x})^2
-m\mu(\vec{B}\times\vec{\x})\cdot(\vec{\omega}\times
\vec{\x})\right],}} where $V(\vec{\x})$ stands for the potential
energy of the particle in static inertial frame,
$-\half\vec{B}\times\vec{\x}$ is the vector potential
$\vec{A}(\vec{\x})$. In the path integral quantization, one should
include the factor $\int {\cal D}\x \exp(iI_{\x}/\hbar)$.

We can quantize \xraction\ canonically. Clearly, the canonical
momentums of $\vec{\x}$ are $\vec{\p}=m\left({d\vec{\x}\over
dt}+\vec{\omega}\times\vec{\x}+\mu\vec{B}\times\vec{\x}\right)$. The
Hamiltonian can be written out \eqn\hHrx{H_{\x}={\vec{\p}^2\over
2m}-(\mu\vec{B}+\vec{\omega})\cdot \vec{L}+V_{eff}(\vec{\x}),} where
$\vec{L}=\vec{\x}\times \vec{\p}$ is the angular momentum operator,
and the effective potential $V_{eff}(\vec{\x})$ is given by
$V_{eff}(\vec{\x})=V(\vec{\x})+\half \mu^2m(\vec{B}\times
\vec{\x})\cdot (\vec{B}\times \vec{\x})$.

\bigskip\noindent{\it{Spin-Orbital Coupling}}

We'll now discuss the coupling between the spin and the spacial
variables by treating it as a perturbation to action
$I_{B\CS}+I_{\x}$. In the static frame, the simplest such term that
preserves the parity is $\xi(\vec{\x})\vec{S}(\phi)\cdot
[\vec{\x}\times({d\vec{\x}\over dt}+e\half \vec{B}_{iner}\times
\vec{\x})]$, where the small factor $\xi(\vec{\x})$ is a some-what
arbitrary function of $\vec{\x}$, for the case of H atom,
$\xi(\vec{\x})={1\over 2mc^2}{1\over
|\vec{\x}|}{dV_{C}(\vec{\x})\over d|\vec{\x}|}$. Now, go to the
$\CS$ frame, we'll have the term $I_{\CS LS}=\int dt
\xi(\vec{\x})[\vec{\x}\times({d\vec{x}\over dt} +\vec{\omega}\times
\vec{\x}+e\half \vec{B}\times \vec{\x})]\cdot \vec{S}(\phi)$. In
terms of path integral quantization, one should insert $\exp(iI_{\CS
LS}/\hbar)$ in $\int \CD\phi\CD\bar{\phi}\CD\lambda\CD\x
\exp\{i{1\over\hbar}(I_{\CS B}+I_{\x})\}$. In terms of canonical
quantization, the spin-orbital coupling is just the usual
perturbation $V_{LS}=\xi(\vec{\x})\vec{L}\cdot \vec{S}$ to the
potential energy of the system.

\newsec{Neutron Interference, And Rabbi Oscillating}

We now apply the general formalism in above section to give unified
interpretations of some well known quantum mechanical effects or
experiments concerning rotating reference of frame.

\bigskip\noindent{\it{Neutron interference}}

In the neutron interference experiment Earth is the rotating frame
$\CS$. The coupling between the angular momentum and rotation of the
frame will course a phase shift $\Delta\varphi$. The relevant terms
of the action is the kinematic terms $I_{\rm k}=\int dt\left[\half
m\left({d\vec{\x}\over dt}\right)^2+i\phi^{\dagger}{d\over
dt}\phi-\lambda(\phi^{\dagger}\phi-n)\right]$ plus the terms account
for the inertial force $I_{\rm r}=\int dt \left[m(\vec{\x}\times
{d\vec{\x}\over dt})+\vec{S}(\phi)\right]\cdot\vec{\omega}$. In the
interference experiment, the insertion of $\exp(iI_{\rm r}/\hbar)$
in the path integration will contribute a phase factor
$\exp(i{\vec{S}\over\hbar}\cdot\oint dt
\vec{\omega})\exp(i{2m\vec{\omega}\over
\hbar}\cdot\half\oint_C(\vec{\x}\times d\vec{\x}))$, where
$\half\oint_C(\vec{\x}\times d\vec{\x})$ is just the area
$\vec{A}_C$ surrounded by the path of the neutron, and $\oint dt
\vec{\omega}$ equals to $2T\vec{\omega}$, where $T$ is the flying
time of the neutron. Thus, the total phase shift
$\Delta\varphi={1\over \hbar}2(m\vec{A}_C+T\vec{S})\cdot
\vec{\omega}$, where the first term is the so called Sagnac effect.
In terms of the Hamiltonian formalism, the relevant terms are
$H_{\rm r}=-\vec{\omega}\cdot(\vec{L}+\vec{S})$, which has been
derived to explain the interference experiment.

\bigskip\noindent{\it{Rabbi Oscillating}}

In the nuclear-magnetism resonance experiment. A magnetic momentum
$g\mu_B$ is coupled to a control magnetic field
$\vec{B}_{\parallel}$ and a transverse rotating field
$\vec{B}^{+}_{\perp}e^{i\omega t}+\vec{B}^{-}_{\perp}e^{-i\omega
t}$. In the frame $\CS$ rotating along the direction of
$\vec{B}_{\parallel}$ with angular velocity $\omega$, this system is
easy to solve. The relevant terms of the action are the kinematic
terms $I_{\CS \rm k}=\int dt \left[i\phi^{\dagger}{d\over
dt}\phi-\lambda(\phi^{\dagger}\phi-n)\right]$, plus the terms
$I_{B\rm r}=\int dt (\vec{\omega}+g\mu\vec{B})\cdot \vec{S}(\phi)$.
At the resonant frequency $\vec{\omega}+g\mu\vec{B}_{\parallel}=0$,
the factor $\exp\{{i\over \hbar}\int dt (g\mu\vec{B}_{\perp}\cdot
\vec{S})\}$ will cause the state of the spin oscillating between the
up and down state with oscillating frequency
$\omega_{R}={g\mu|\vec{B}_{\perp}|\over 2\hbar}$. In the more
general case the oscillating frequency $\Omega$ is given by
$\Omega^2=\left({\vec{\omega}+g\mu\vec{B}_{\parallel}\over
2\hbar}\right)^2+\omega_R^2$. All these results can also be derived
from the Hamiltonian $H_{B\CS}=-g\mu\vec{B}\cdot
\vec{S}-\vec{\omega}\cdot \vec{S}$.

\bigskip\noindent{\it{Equation Of Motions}}

We now digress to discuss the equation of motions--in the rotating
frame $\CS$--derived by using Schwinger's quantum action principle
to see the quantum extension of the ordinary non-inertial force in
the classical theory, e.g. the Coriolis force.

The relevant action is $I=\int dt\left[\half m\left({d\vec{\x}\over
dt}\right)^2+m\vec{\omega}\cdot(\vec{\x}\times {d\vec{\x}\over
dt})-V(\vec{\x})+\half m(\vec{\omega}\times\vec{\x})^2\right]$. The
quantum action principle tells us $\langle\psi_f|\delta
I|\psi_i\rangle=0$. Carrying out the variation of $I$ will give us
the equation of motion $\langle \psi_f|m{d^2\vec{\x}\over
dt^2}+2m{d\vec{\x}\over dt}\times
\vec{\omega}+m(\vec{\omega}\times\vec{\x})\times\vec{\omega}|\psi_i\rangle=-\langle\psi_f|{\pt
V(\vec{\x})\over \pt\vec{\x}}|\psi_i\rangle$. which is just the
quantum mechanical generalization of the ordinary Newton equation in
rotating frame, where the Coriolis force $2m\vec{\omega}\times
{d\vec{\x}\over dt}$ comes from the $\int dt
m\vec{\omega}\cdot(\vec{\x}\times {d\vec{\x}\over dt})$ term of the
action $I$. In terms of the Hamiltonian formalism. The Coriolis
force comes from the $-\vec{\omega}\cdot \vec{L}$ term of the
Hamiltonian $H={\vec{\p}^2\over 2m}-\vec{\omega}\cdot
\vec{L}+V(\vec{\x})$.

\listrefs
\end